\documentclass[prd,superscriptaddress,twocolumn,showpacs,nofootinbib]{revtex4}
\usepackage{amsfonts,amsmath,amssymb}
\usepackage{graphicx}

\newcommand{\Tc}{T_\mathrm{c}}
\newcommand{\Tdiss}{T_\mathrm{diss}}
\newcommand{\feyn}[1]{
  \setbox0=\hbox{\ensuremath{#1}}
  \hbox to\wd0{\hbox to0pt{\hbox to\wd0{\hss/\hss}\hss}\box0}}
\newcommand{\EF}{E_\mathrm{F}}
\newcommand{\pF}{p_\mathrm{F}}
\newcommand{\NF}{N_\mathrm{F}}
\newcommand{\NFbar}{N_\mathrm{\bar F}}
\newcommand{\NB}{N_\mathrm{B}}
\newcommand{\NBbar}{N_\mathrm{\bar B}}
\newcommand{\Ntotal}{N_\mathrm{total}}
\newcommand{\MB}{M_\mathrm{B}}
\newcommand{\Ep}{E_{\boldsymbol p}}

\def\Jvol<#1,#2,#3>{#1}
\def\Jpage<#1,#2,#3>{#2}
\def\Jyear<#1,#2,#3>{#3}
\newcommand\journal[1]{\textbf{\Jvol<#1>}, \Jpage<#1> (\Jyear<#1>)}
\newcommand\PRL[1]{Phys.\ Rev.\ Lett.\ \journal{#1}}
\newcommand\PRA[1]{Phys.\ Rev.\ A \journal{#1}}
\newcommand\PRC[1]{Phys.\ Rev.\ C \journal{#1}}
\newcommand\PRD[1]{Phys.\ Rev.\ D \journal{#1}}

\newcommand\PTP[1]{Prog.\ Theor.\ Phys.\ \journal{#1}}

\newcommand\JLTP[1]{J.\ Low Temp. Phys.\ \journal{#1}}

\newcommand\IJMPA[1]{Int.\ J.\ Mod.\ Phys.\ A \journal{#1}}

\begin{document}

\title{BCS-BEC crossover in a relativistic superfluid
       and its significance to quark matter}

\author{Yusuke~Nishida}
\email[E-mail:~]{nishida@nt.phys.s.u-tokyo.ac.jp}
\affiliation{Department of Physics, University of Tokyo,
             Tokyo 113-0033, Japan}
\author{Hiroaki~Abuki}
\email[E-mail:~]{abuki@yukawa.kyoto-u.ac.jp}
\affiliation{Yukawa Institute for Theoretical Physics, Kyoto University, 
             Kyoto 606-8502, Japan}

\date{\today}

\begin{abstract} 
The character change of a superfluid state due to the variation of the
 attractive force is investigated in the relativistic framework with
 a massive fermion. 
Two crossovers are found. 
One is a crossover from the usual BCS state to the Bose-Einstein
 condensation (BEC) of bound fermion pairs. 
The other is from the BEC to the relativistic Bose-Einstein condensation 
 (RBEC) of nearly massless bound pairs where 
 antiparticles as well as particles dominate the thermodynamics.
Possible realization of the BEC and RBEC states in the quark matter is
 also pointed out.
\end{abstract}

\maketitle

Recently, new superfluid states in the ultracold gas of fermionic 
 alkali atoms $(^{40}\mathrm{K}, ^{6}\mathrm{Li})$ were
 realized \cite{regal04}. 
Using the Feshbach resonance, the long-standing idea of the
 crossover from the BCS state to the Bose-Einstein condensation (BEC)
 \cite{eagles69,leggett80,nozieres85} has been extensively examined. 
The basic concept of the BCS-BEC crossover is as follows: As long as the 
 attractive interaction between fermions is weak, the system exhibits 
 the superfluidity characterized by the energy gap in the BCS
 mechanism. 
On the other hand, if the attractive interaction is strong enough,
 the fermions first form bound molecules (bosons). 
Then they start to condense into the bosonic zero-mode at some 
 critical temperature.
These two situations are smoothly connected without the phase
 transition. 

The possible realization of the BCS-BEC crossover in various
 systems has been theoretically investigated.
These include the liquid $^3$He \cite{leggett80}, the trapped alkali
 atoms \cite{ohashi02}, and the nuclear matter \cite{stein95}. 
One of the most striking features of the crossover is that the critical
 temperature in the BEC region is independent of the coupling for 
 the attraction between fermions. 
This is because the increase of the coupling only affects the internal
 structure of the bosons, while the critical temperature is determined
 by the boson's kinetic energy. 
Thus, the critical temperature reaches a ceiling for the large coupling 
 as long as the binding effect on the boson mass can be neglected.
Even in the nuclear matter where the interaction is relatively strong,
 the binding energy of the deuteron is much smaller than the nucleon
 mass. 
This fact allows us to work within a nonrelativistic framework for
 describing such a crossover. 

It is interesting to ask how the situation changes in relativistic
 systems where the binding effect can not be neglected.
The color superconducting phase in the dense quark matter
 \cite{rajagopal01,abuki02} and the pion superfluid phase at finite
 isospin density \cite{son00} would be examples.
In this article, we will show that there could be two crossovers in the
 relativistic superfluids. 
One is the ordinary BCS-BEC crossover, where the critical temperature in 
 the BEC region would not plateau because of the relativistic effect. 
The other is from the BEC state to the novel state, 
 {\em the relativistic BEC (RBEC)}, where the critical temperature
 increases to the order of the Fermi energy.

In order to explore the BCS-BEC and BEC-RBEC crossovers in the
 relativistic system, we start with the following contact four-Fermi
 interaction model: 
\begin{align}
\begin{split}
 \mathcal{L}[\psi,\bar\psi]
 &=\bar\psi\left(i\feyn\partial-m+\gamma_0\mu\right)\psi\\
 &\qquad+G\left(i\psi^\dagger\gamma_5C
 \psi^*\right)
 \cdot\left(i\psi^\mathrm{T}\gamma_5C
 \psi\right).\label{eq:lagrangian}
\end{split}
\end{align}
Here $\psi=\psi(t,{\boldsymbol x})$ is a
Dirac fermion field having a spinor index implicitly. 
The fermion mass and chemical potential are denoted by $m$ and
 $\mu$. 
$C=i\gamma_0\gamma_2$ is the charge conjugation matrix and 
$G$ is a coupling constant for the attraction in the $J^P=0^+$ channel.
Qualitative results shown below are not modified even when fermions have
 internal degrees of freedom other than spin. 
Therefore, we will make our analysis without them for simplicity.

The partition function can be written as
\begin{align}
 Z=\int\mathcal D\psi\mathcal D\bar\psi\,
 \exp\left(-\int_0^{1/T}d\tau\int d{\boldsymbol x}\,
 \mathcal L_\mathrm E[\psi,\bar\psi]\right),
\end{align}
 where $\mathcal L_\mathrm E$ is the Lagrangian density in the Euclidean 
 space. 
Introducing Hubbard-Stratonovich fields
 $\Delta(\tau,{\boldsymbol x})$ for $i\psi^\mathrm{T}\gamma_5C\psi$ and
 $\Delta^{\!*}(\tau,{\boldsymbol x})$ for $i\psi^\dagger\gamma_5C\psi^*$
 and integrating out the fermion fields lead to
\begin{align}
 Z=Z_0\int\mathcal D\Delta\mathcal D\Delta^{\!*}\,
 \exp\left(-S_\mathrm{eff}[\Delta,\Delta^{\!*}]\right).
 \label{eq:partition function}
\end{align}
Here $Z_0=e^{-\beta\Omega_0(\mu,T)}$ is the free fermion part of 
 the partition function, while $S_\mathrm{eff}[\Delta,\Delta^{\!*}]$ is
 the effective action for the bosonic fields. 
In order to include the effect of the fluctuation, 
 we evaluate the functional integral in the Gaussian approximation,
 whose validity will be discussed later. 
Expansion of the effective action up to the second order 
 in $\Delta$ results in
\begin{align}
\begin{split}
 S_\mathrm{eff}[\Delta,\Delta^{\!*}]
 \simeq T\sum_n\int\frac{d{\boldsymbol p}}{(2\pi)^3}
 \left[\frac1{G}-\chi(i\omega_n,{\boldsymbol p})\right]
 \bigl|\tilde\Delta(i\omega_n,{\boldsymbol p})\bigr|^2,
 \label{eq:expansion} 
\end{split} 
\end{align}
 where $\tilde\Delta(i\omega_n,{\boldsymbol p})$ is the momentum
 representation of the pair field with $\omega_n=2\pi nT$ being the
 Matsubara frequency. 
$\chi(i\omega_n,{\boldsymbol p})$ is a pair susceptibility at the one
 loop level \cite{kitazawa02}. 
The critical temperature for the superfluidity, $\Tc$, is given by the
 solution of the equation:
\begin{align}
 \frac1G-\chi(0,{\boldsymbol 0})\Big|_{T=\Tc}=0.
 \label{eq:thouless}
\end{align}
This is nothing but the Thouless criterion which states that 
 the pair fluctuation becomes tachyonic at low momentum because of
 $1/G-\chi(0,{\boldsymbol 0})|_{T<\Tc}<0$. 
This is the signal of the BCS instability to the formation of Cooper
 pairs. 

The integration over $\Delta, \Delta^{\!*}$ in
 Eq.~(\ref{eq:partition function}) leads to the thermodynamic potential
 in the Gaussian approximation:
 $\Omega(\mu,T)=\Omega_0(\mu,T)+\Omega_\mathrm{fluc}(\mu,T)$ with 
\begin{align}
 \Omega_\mathrm{fluc}(\mu,T)=T\,\sum_n
 \int\frac{d{\boldsymbol p}}{(2\pi)^3}
 \log\left[\frac1G-\chi(i\omega_n,{\boldsymbol p})\right].
\end{align}
Following Nozi$\mathrm{\grave e}$res and Schmitt-Rink \cite{nozieres85},
 $\Omega_\mathrm{fluc}$ can be written in terms of a phase shift
 $\delta(\omega,{\boldsymbol p})$ defined by 
 $\delta(\omega,{\boldsymbol p})
 =-\arg\left[1/G-\chi(\omega+i0,{\boldsymbol p})\right]$.
By differentiating the thermodynamic potential with $\mu$, we obtain the
 fermion number density as follows: 
\begin{align}
\begin{split}
 N_\mathrm{total}&=2\int\frac{d{\boldsymbol p}}{(2\pi)^3}
 \left\{f_{\mathrm F}(\Ep-\mu)-f_{\mathrm F}(\Ep+\mu)\right\}\\
 &\quad+\int\frac{d{\boldsymbol p}}{(2\pi)^3} 
 \int_{-\infty}^\infty\frac{d\omega}\pi\,f_{\mathrm B}(\omega)\,
 \frac{\partial\delta}{\partial\mu}(\omega,{\boldsymbol p})
 \label{eq:density}
\end{split}
\end{align}
 with $f_{\mathrm F}(\omega)=1/(\exp[\omega/T]+1)$ being the Fermi
 distribution function and $f_{\mathrm B}(\omega)
 =\mathrm{sign}(\omega)/\left(\exp[\left|\omega\right|/T]-1\right)$
 being the Bose distribution function.%
\footnote{We use this Bose distribution function so that the
 thermodynamic potential is symmetric under $\mu\to-\mu$.}
The first term which we denote by $N_\mathrm{MF}=\NF-\NFbar$ represents
 the contribution of fermions and antifermions at the mean field level
 and the second one which we denote by $N_\mathrm{fluc}$ represents the
 contribution of pair fluctuations.
Instead of $\Ntotal$, we will sometimes use the Fermi momentum $\pF$,
 which is defined by $\Ntotal=\pF^{\,3}/3\pi^2$. 

If the attraction is strong enough, bound states appear and 
 we can extract the bound boson (antiboson) contribution 
 $\NB$ $(\NBbar)$ from $N_\mathrm{fluc}$ \cite{nozieres85,ohashi02}.
By picking up the bound state poles in
 $\partial\delta(\omega,{\boldsymbol p})/\partial\mu$
 in the $\omega$-integral of Eq.~(\ref{eq:density}), we obtain
\begin{align}
 \NB&=\int\frac{d{\boldsymbol p}}{(2\pi)^3}
 \left[2-\frac{\partial\omega_\mathrm B({\boldsymbol p})}
 {\partial\mu}\right]
 f_{\mathrm B}\left(\omega_\mathrm B({\boldsymbol p})-2\mu\right)
\intertext{and}
 \NBbar&=\int\frac{d{\boldsymbol p}}{(2\pi)^3}
 \left[2+\frac{\partial\omega_\mathrm{\bar B}({\boldsymbol p})}
 {\partial\mu}\right]
 f_{\mathrm B}\left(\omega_\mathrm{\bar B}({\boldsymbol p})+2\mu\right).
 \label{eq:anti-boson density}
\end{align}
Here, $\omega_\mathrm B({\boldsymbol p})$ and 
 $-\omega_\mathrm{\bar B}({\boldsymbol p})$ are the solutions of 
 $1/G-\chi(\omega-2\mu,{\boldsymbol p})=0$ and correspond to the
 energy of the boson and the antiboson, respectively. 
Then the remaining part
 $N^\mathrm{unstable}=N_\mathrm{fluc}-(\NB-\NBbar)$ can be 
 interpreted as the contribution of unstable off-shell bosons.

\begin{figure}[tp]
  \includegraphics[width=0.45\textwidth,clip]{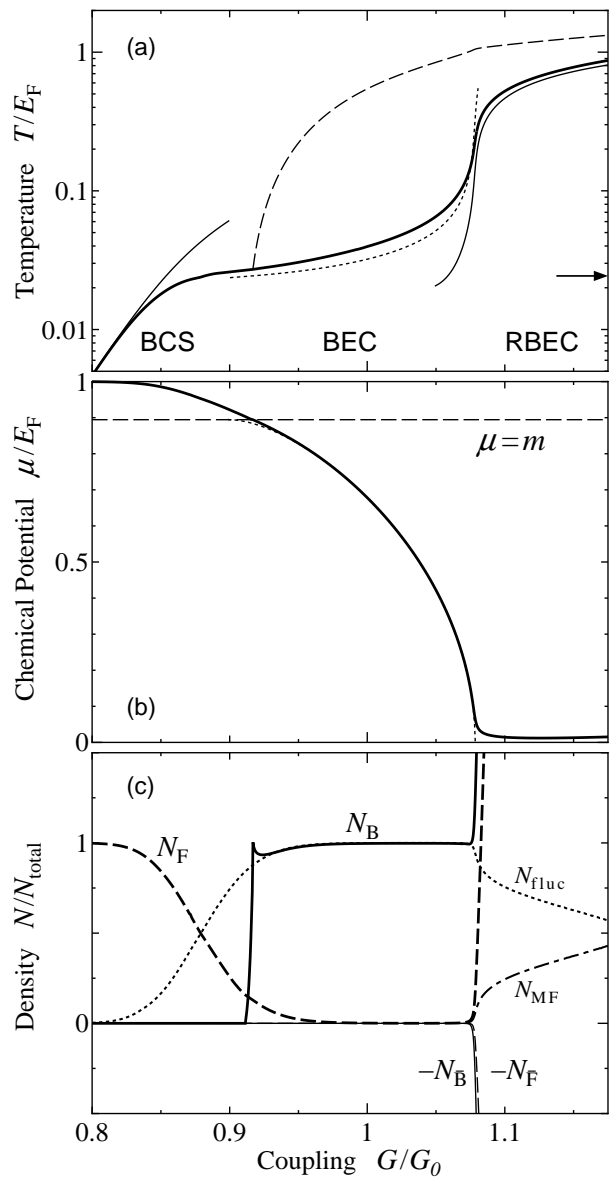}
  \caption{(a) Critical temperature $\Tc$ normalized by the Fermi energy
  $\EF=\sqrt{m^2+\pF^2}$ (thick solid line) as a function of the
  coupling $G/G_0$. $G_0$ is defined by $G_0=\pi^2/\Lambda^2$. 
  For other lines, see the text. 
  \,(b) Chemical potential $\mu/\EF$ as a function of the coupling
  $G/G_0$. The dashed line represents the level where $\mu=m$;
  $m/\EF\simeq0.89$ in the present case. The dotted line corresponds to 
  one half of the bound boson's mass in the vacuum. 
  \,(c) The ratios of $\NF,\NFbar,\NB$ and $\NBbar$ to the fixed
  $\Ntotal$ as functions of $G/G_0$. 
  $N_\mathrm{MF}/\Ntotal$ and $N_\mathrm{fluc}/\Ntotal$ are also
  plotted. The line for $N_\mathrm{MF}$ is behind that for $\NF$ for 
  $G/G_0\lesssim1.07$ because of the absence of antifermions.
  \label{fig}}\vspace{-3mm}
\end{figure}

In numerical calculations, a momentum cutoff $\Lambda$ is introduced 
 in order to regularize the ultraviolet divergence and all the
 dimensionful quantities are scaled by $\Lambda$.
We take a characteristic parameter set ($m/\Lambda=0.2$ and
 $\pF/\Lambda=0.1$) so that we can analyze the effect of relativity. 
We confirmed that the variation of $\pF$ does not change our 
 qualitative arguments below. 
Also, how the variation of $m$ affects our results will be discussed
 later. 
Figures \ref{fig}(a) and \ref{fig}(b) show numerical results of the
 critical temperature $\Tc$ and the chemical potential $\mu$ as
 functions of $G$ with the total number density $\Ntotal$ fixed, which
 are obtained by solving Eqs.~(\ref{eq:thouless}) and (\ref{eq:density}) 
 simultaneously. 
The ratios of the fermion and stable boson densities to $\Ntotal$
 at $T=\Tc$ are also plotted in Fig.~\ref{fig}(c).
Based on these three figures, we will argue that there are three
 physically distinct regions; the weak, intermediate, and strong coupling
 regions. 
The superfluid states realized in the three regions will be interpreted 
 as the BCS, BEC, and relativistic BEC phases, respectively.

In the weak coupling region $G/G_0\lesssim0.86$, $\Tc$ increases
 exponentially as is well-known in the weak coupling BCS theory. 
Its behavior is well described by the mean field approximation 
 (the left thin solid line in Fig.~\ref{fig}(a)). 
$\mu$ in this region is almost equal to the Fermi
 energy $\EF=\sqrt{m^2+\pF^{\,2}}$.
Accordingly, the fermion density $\NF$ dominates the total density.
From these facts, the superfluid state realized in this region can be
 regarded as the BCS state.

In the intermediate coupling region $0.86\lesssim G/G_0\lesssim1.07$,
 $\Tc$ increases much slowly and $\mu$ decreases monotonously. 
Once $\mu$ becomes smaller than $m$, stable bosons with the
 mass $\MB(\Tc)=2\mu$ appear and they dominate the total density.%
\footnote{The apparent singularity in the stable boson density in
 Fig.~\ref{fig}(c) does not mean a phase transition. The total boson
 density $N_\mathrm{fluc}$ is a smooth and positive function of the
 coupling.}
The critical temperature for the ideal Bose gas is approximately given
 by 
\begin{align}
 T_\mathrm{BEC}^\mathrm{NR}=\frac{2\pi}{\MB}
 \left[\frac{\NB}{2\zeta(3/2)}\right]^{2/3}.
\end{align}
This nonrelativistic formula for the BEC critical temperature with the
 boson mass $\MB(\Tc)=2\mu$ is examined by the dotted line in
 Fig.~\ref{fig}(a), which well approximates 
 $\Tc$ in the intermediate coupling region.
Therefore, we can interpret the superfluid state realized in this region 
 is in the BEC phase.
The slowly increasing $\Tc$ in the BEC phase is in contrast to the
 result of the nonrelativistic calculation where the critical
 temperature approaches to a constant value \cite{melo93}. 
In the nonrelativistic framework, the change of the boson mass due to
 the binding can be neglected by definition.
Thus, the critical temperature in the BEC phase is given by 
 $T_\mathrm{BEC}^\mathrm{NR}$ with $\MB=2m$, which is indicated by the
 arrow in Fig.~\ref{fig}(a), and is independent of the coupling. 
In our relativistic framework, however, the boson mass $\MB(\Tc)=2\mu$
 can become smaller, and consequently, 
 $\Tc$ becomes larger as one increases the coupling. 

We remark that the criterion for the BEC state, $\mu<m$, is model
 independent. 
At $T=\Tc$ of the BEC, the chemical potential for the boson
 $\mu_\mathrm{B}=2\mu$ should be equal to its mass $\MB$. 
On the other hand, $\MB$ must be less than $2m$ for the binding. 
Therefore, we have $\mu<m$ in the BEC region. 

Let us discuss the strong coupling region $G/G_0\gtrsim1.07$, where 
 $\Tc$ rapidly increases and the nonrelativistic formula for the BEC
 critical temperature breaks down. 
Because $\mu$ is smaller than $\Tc$ in this region, 
 antiparticles can be easily excited. 
As is shown in Fig.~\ref{fig}(c), the antifermion and antiboson
 densities $(\NFbar, \NBbar)$ grow rapidly. 
At the same time, the fermion and stable boson densities 
 $(\NF, \NB)$ increase so that the total number density
 is unchanged.
$\Tc$ in this region can be approximated by the
 ideal BEC critical temperature in the relativistic limit \cite{kapsta}:
\begin{align}
 T_\mathrm{BEC}^\mathrm{RL}=\sqrt{\frac{3(\NB-\NBbar)}{2\MB}}. 
\end{align}
We note this approximate formula slightly deviates from $\Tc$ 
 particularly in the large coupling region. This is because a large
 number of fermions are accompanied there, which is favorable in terms
 of entropy.
We refer to the boson condensed phase with antiparticles in the
 strong coupling region as the relativistic BEC (RBEC) phase.

We can also understand the rapid increase of $\Tc$ in terms of the
 decreasing mean interparticle distance $\bar d_\mathrm p$. 
Let us estimate the critical temperature in the RBEC phase by comparing 
 the thermal de Broglie wavelength $\pi/(\sqrt3\,T)$ to 
 $\bar d_\mathrm p$.
We estimate $\bar d_\mathrm p$ by $\NF^{-1/3}$ since fermions give a
 dominant contribution to the density in the strong coupling region. 
Thus, we have $T\sim\pi\NF^{1/3}/\sqrt3$, which agrees with $\Tc$ well 
 (see the right thin solid line in Fig.~\ref{fig}(a)). 

An essential difference between the BCS and (R)BEC phases is
 that the stable bosons are present above $\Tc$ in the (R)BEC region.
As the temperature is increased, the binding energy of the stable boson 
 decreases. 
The bound state pole eventually disappears at a certain temperature,
 which we call a dissociation temperature $\Tdiss$. 
$\Tdiss$ as a function of the coupling is shown in Fig.~\ref{fig}(a)
 by the dashed line. 
$\Tdiss$ line appears from the point $G/G_0\simeq0.92$ where the fermion
 pairs start to form the bound bosons. 
They get bound deeper with the increasing coupling,
 and as a consequence, $\Tdiss$ increases monotonically. 
$\Tdiss$ line separates the normal phase into two regions; 
 a normal phase without stable bosons for $T>\Tdiss$, and a 
 {\em preformed boson phase} with stable bosons for $\Tc<T<\Tdiss$. 
The preformed boson phase in the intermediate and strong coupling
 regions may provide us with new insight into the precursory phenomenon
 above $\Tc$ \cite{kitazawa02}. 

We have discussed the character change of the superfluid state with a
 specific set of $m$ and $\pF$.
Let us now discuss the $m$-dependence of the crossover boundaries with
 keeping $\pF$ fixed. 
The crossover boundary from the BCS to the BEC is characterized by the
 point where the bound states are formed in the medium. 
On the other hand, the crossover boundary from the BEC to the RBEC is
 characterized by the point where the boson mass $2\mu$
 becomes smaller than $\Tc$.
These two points are well approximated by the coupling $G_1$ where
 the boson is formed and $G_2$ where it becomes massless in the
 vacuum with $\mu=T=0$ (see the dotted line in Fig.~\ref{fig}(b) which
 is indistinguishable from the solid line). 
We can show that $G_{1(2)}$ increases (decreases) as decreasing the
 fermion mass $m$.
It means that it becomes hard to bind two fermions due to the larger
 kinetic energy for a smaller $m$, while less attraction will be needed
 to cancel $2m$ by the binding.
Accordingly, the BEC region shrinks with decreasing $m$, while the RBEC
 dominates the larger region in the coupling space.
In the ultrarelativistic limit $m\to0$, the BEC region will disappear
 because fermions could no longer be bound.
Even in this case, we still have a superfluid phase with $2\mu<\Tc$ for
 the large coupling which is smoothly connected with the RBEC phase at
 $m>0$.

In summary, we have discussed two crossovers in the relativistic
 four-Fermi model with the massive fermion: 
One is the crossover from the usual BCS to the BEC of bound fermion
 pairs and the other is that from the BEC to the RBEC of nearly massless
 bound pairs. 
In order to avoid the cutoff artifacts, we have checked each of 
 $\NF, \NB, \NFbar$ and $\NBbar$ does not exceed $(\Lambda/2\pi)^3$
 within the coupling range shown here. 

Since we have employed the Gaussian approximation which corresponds to
 the resummation of ring diagrams into the thermodynamic potential,
 all the interactions among bosons are neglected in our analysis. 
However, it is shown in the nonrelativistic framework that
 the 2-body interaction between bosons becomes smaller with increasing
 the attraction between fermions \cite{melo93}. 
The multibody scatterings among bosons are also negligible in a dilute gas. 
Thus, our approximation is valid except for the vicinity of the
 BCS-BEC crossover boundary and the very dense RBEC limit.
Going beyond the Gaussian approximation so as to take into account the
 interactions among bosons in our relativistic framework is an
 important future issue. 
Other approaches to the BCS-BEC crossover also may be useful
 \cite{babaev01}. 

Finally, we make some speculative remarks on the relevance of the phases 
 discussed above to QCD. 
The BCS-BEC crossover, which takes place for $\mu\gg T$, may be realized
 in the cold dense quark matter \cite{abuki02}. 
The fermion mass $m$ in Eq.~(\ref{eq:lagrangian}) in this case 
 should be interpreted as the current or dynamical quark masses. 
Also, it is an interesting future problem to generalize
 our model by taking into account the plasmino mass 
 $m_\mathrm{pl}\sim g\sqrt{\mu^2+\pi^2T^2}$ with $g$ being the QCD
 coupling constant. 
The plasmino mass can play a role of the chiral invariant mass
 constituting the boson mass. 
Further study with the plasmino effective action \cite{braaten92}
will give us more insight into the realistic
 BCS-BEC crossover in the quark matter. 
In fact, the BEC criterion $\mu<m_\mathrm{pl}$ leads to $g\gtrsim1$ for
 $\mu\gg T$, which corresponds to the density relevant to the center of
 compact stars. 

The BEC-RBEC crossover, which takes place for $T\gg\mu\sim0$, will
 be relevant to the quark-gluon plasma just above the deconfinement
 transition. 
Possibility of having not only $q\bar{q}$ bound states
 \cite{hatsuda85} but also $qq$ bound states \cite{shuryak04} in
 the deconfined phase has close relevance to the bound bosons in the
 RBEC state in this article. 
The plasmino mass $m_\mathrm{pl}\sim gT$ again will play a crucial
 role to have bound bosons in the realistic situation. 

The above discussions suggest that there is a band of superfluid
 phases (BCS-BEC-RBEC) between the hadronic phase and the quark-gluon
 plasma phase in the QCD phase diagram. 
Also, the preformed boson phase may exist between the (R)BEC phase and 
 the quark-gluon plasma phase. 
The transport properties in these phases are also of great interest. 
In the nonrelativistic framework, it is  shown that the pair
 fluctuation is dissipative in the BCS region, while it propagates
 without viscous damping in the BEC region \cite{melo93}. 
The detailed analysis including all the relevant hydrodynamic modes
 may provide a picture for almost the perfect fluid aspect of
 quark-gluon plasma. 

The generalization of our work so as to allow for the $q\bar{q}$
 condensation will be essential to see the realization of the BEC and
 RBEC states in QCD. 
It is known that the one-gluon exchange generates the attraction in the
 scalar $q\bar{q}$ channel whose strength is 2 times larger than that
 in the $qq$ channel. 
Even though the $qq$ condensation still has a kinematic advantage due
 to the existence of the Fermi surface, the dominant $q\bar{q}$
 attraction may wash out the (R)BEC phase leading to the large
 $q\bar{q}$ condensation. 
Whether the (R)BEC phase survives in the QCD phase diagram should be
 settled after taking into account the possibility of the $q\bar{q}$
 condensation in our analysis.

\begin{acknowledgments}
 The authors would like to thank T.~Hatsuda for discussions, comments
 and reading manuscript. 
 Y.~N. is supported by the Japan Society for the
 Promotion of Science for Young Scientists. 
 H.~A. is supported by the 21COE program ``Center for Diversity and
 Universality in Physics'' at Kyoto University.
\end{acknowledgments}

\end{document}